# Coupling of microwave magnetic dynamics in thin ferromagnetic films to stripline transducers in the geometry of the broadband stripline ferromagnetic resonance

*M. Kostylev**

*School of Physics, the University of Western Australia, Crawley 6009, Australia*

Abstract. We constructed a quasi-analytical self-consistent model of the stripline-based broadband ferromagnetic resonance (FMR) measurements of ferromagnetic films. Exchange-free description of magnetization dynamics in the films allowed us to obtain simple analytical expressions. They enable quick and efficient numerical simulations of the dynamics. With this model we studied the contribution of radiation losses to the ferromagnetic resonance linewidth, as measured with the stripline FMR. We found that for films with large conductivity of metals the radiation losses are significantly smaller than for magneto-insulating films. Excitation of microwave eddy currents in these materials contributes to the total microwave impedance of the system. This leads to impedance mismatch with the film environment resulting in decoupling of the film from the environment and, ultimately, to smaller radiation losses. We also show that the radiation losses drop with an increase in the stripline width and when the sample is lifted up from the stripline surface. Hence, in order to eliminate this measurement artefact one needs to use wide striplines and introduce a spacer between the film and the sample surface. The radiation losses contribution is larger for thicker films.

## 1. Introduction

The microwave conductivity contribution to the stripline broadband ferromagnetic resonance (FMR) response of highly-conducting (metallic) magnetic multilayers and nanostructures of sub-skin-depth thicknesses has attracted significant attention in recent years [1-17]. It has been shown that these effects are important when the microwave magnetic field is incident on only one of the two surfaces of a planar metallic material (see e.g. [17]).

The geometry of a stripline ferromagnetic resonance experiment [17-21] is characterized by such single-surface incidence of the microwave magnetic field on the sample. This experiment usually employs a macroscopic coplanar (CPW) or microstrip (MSL) stripline through which a microwave current flows (Fig. 1). A sample - a film or a nanostructure - sits on top of this line, often separated by an insulating spacer. The stripline with the sample is placed in a static magnetic field applied along the stripline. The microwave current in the stripline drives magnetization precession in the ferromagnetic material. The complex transmission coefficient S21 of the stripline is measured either as a function of microwave frequency $f$ for a given applied magnetic field $\mathbf{H}=\mathbf{e}_z H$ ("frequency resolved FMR"), or as a function of $H$ for given $f$ ("field-resolved FMR") to produce FMR traces. The FMR absorption by the material is seen as a deep in the Re(S21) vs. $H$ or $f$ trace.

*mikhail.kostylev@uwa.edu.au

In our previous work [16] we constructed a quasi-analytical theory of stripline broadband FMR of single-layer metallic ferromagnetic films. A drawback of the constructed theory is that it is not self-consistent. It uses the same approach as previously employed to calculate the impedance of microstip transducers of travelling spin waves in thick magneto-insulating films [22,23]. The central point of this approach is assumption of some realistic distribution for the microwave current density across the width of a microstrip ("Given Current Density" (GCD) method). The next step of the solution is calculation of the amplitude of the dynamic magnetization in the film driven by the Oersted field of the assumed microwave current density. Finally, the microwave electric field induced in the stripline by the found dynamic magnetization in the film is determined. This 3-step analysis allows one to obtain the value of the complex impedance inserted into the microwave path due to loading of a section of the microstrip line by the ferromagnetic film.

Later on a self-consistent approach to calculation of the inserted impedance was suggested [24,25]. In the framework of the self-consistent approach the distribution of the microwave current density is obtained by solving an integral equation. Then the found distribution is used to calculate the impedance with one of the same GCD theories [22,23].

In this work we use a similar approach of an integral equation to obtain a self-consistent solution for the broadband stripline FMR of highly conducting ferromagnetic films with nanometre-range thicknesses. To simplify the problem, contrary to [16], we neglect the exchange interaction. In this way we are able to treat the fundamental (dipole) mode of FMR response of thin magnetic films only; responses of the higher-order standing spin wave modes across the film thickness cannot be obtained with this theory. Given the importance of the fundamental mode for various applications of FMR [19,26], this does not represent a major drawback. Furthermore, simple analytical description in the Fourier space which follows from the exchange-free approximation results in an efficient and quick numerical algorithm for solution of the integral equation for the microwave current density in the stripline.

In our discussion we will focus on the effect of coupling of the magnetization dynamics in the film to the microwave current in the stripline. Inclusion of the conductivity effect will allow us to judge whether the conductivity may influence the strength of this coupling. Experimentally, the problem of coupling of the probing stripline to the FMR in a film has been addressed in a recent paper [27]. It has been shown that strong coupling leads to additional resonance linewidth broadening called "radiation damping". This damping mechanism is related to radiation of the magnetization precession energy back into the probing system – the stripline, because of non-negligible coupling between the two. Basically, one deals with the fact that the external and unloaded Q-factors of a resonator are different [28] if coupling of the resonator to environment is not vanishing. The theory developed in the present work includes naturally the radiation damping, as well as damping due to eddy currents and excitation of travelling spin waves.

## 2. Numerical model

To solve the problem we make use of the idea first proposed in [9]. We extend it to the case of electromagnetic boundary conditions appropriate for excitation of magnetization dynamics in a ferromagnetic film by a stripline [16]. These boundary conditions include microwave shielding effect by the eddy currents in conducting films [10,11].

We consider a model in which the *y*-axis is perpendicular to the surfaces of a conducting magnetic film (Fig. 1(a)). For small elevations *s* of the film from the stripline surface (*s*<<*w*) the microwave field of a microstrip line is localized within the area of the width on the order 2*w*. This allows us to consider the sample size in the direction *x* as infinite.

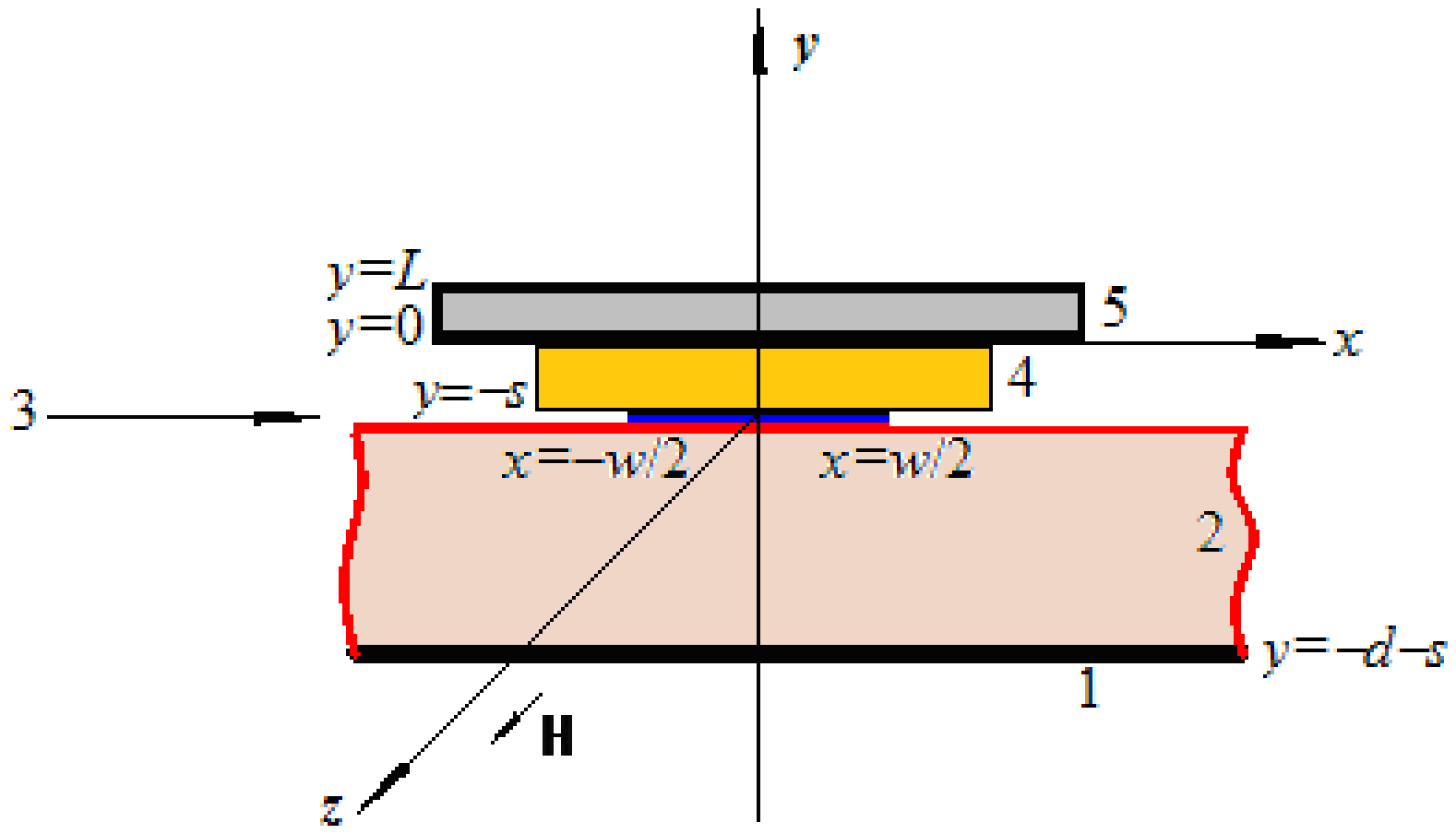

Fig.1. (a) Sketch of the modelled geometry. 1: ground plane of the microstrip line. 2: substrate of the microstrip line of thickness *d*. 3: infinitely thin strip of width *w* carrying a microwave current. 4: Spacer of thickness *s*. 5: ferromagnetic film of thickness *L*.

Thus, the film of thickness *L* is assumed to be continuous in the *x*- and *z*-directions. The external magnetic field $\mathbf{H} = H\mathbf{u}_z$ is applied in the positive direction of the *z*-axis. All the dynamic variables depend harmonically on the time - $\exp(i\omega t)$, where $\omega$ is the microwave frequency. In order to include the magnetization dynamics in the ferromagnetic layer into the model, we employ the linearized Landau-Lifshitz equation

$$i\omega t\mathbf{m} = -|\gamma|(\mathbf{m}\times\mathbf{H} + M_0\mathbf{u}_z\times\mathbf{h}_{eff}). \quad (1)$$

In (1) the dynamic magnetization vector **m** has only two non-vanishing components ($m_x$ , $m_y$) that are perpendicular to the static magnetization $M_0\mathbf{u}_z$, where $M_0$ is the saturation magnetization for the ferromagnetic film, and $\mathbf{u}_z$ is a unit-vector in the direction *z*. The dynamic effective field $\mathbf{h}_{\text{eff}}$ has only one component - the dynamic magnetic field **h** which includes the magnetostatic field of the dynamic magnetization and the Oersted field of the eddy currents which may circulate in the film because of non-vanishing film conductivity $\sigma$.

In this approximation, the Linearized Landau-Lifshits equation reduces to the Polder Microwave Susceptibility Tensor $\hat{\chi}$:

$$\mathbf{m} = \hat{\chi}\mathbf{h} \quad (2)$$

with

$$\hat{\chi} = \begin{vmatrix} \chi & i\chi_a \\ -i\chi_a & \chi \end{vmatrix},$$

where

$$\chi = \frac{\omega_H \omega_M}{\omega_H^2 - \omega^2},\ \chi_a = \frac{\omega\omega_M}{\omega_H^2 - \omega^2},$$

$\omega_M = \gamma M_0$, $\omega_H = \gamma H + i\alpha_G \omega$, $\gamma$ is the gyromagnetic ratio, $\alpha_G$ is the Gilbert magnetic damping constant, and *i* is the imaginary unit. Note that the second term in the last expression accounts for magnetic losses in the system. The losses are included by allowing the applied field to take complex values ($H + i\alpha_G \omega / \gamma$) [29].

The dynamic magnetic field **h** is sought as solution of Maxwell Equations

$$\nabla \times \mathbf{h} = \sigma \mathbf{e}, \quad (3)$$

$$\nabla \times \mathbf{e} = -i\omega\mu_0 (\mathbf{h} + \mathbf{m}), \quad (4)$$

$$\nabla \mathbf{h} = -\nabla \mathbf{m}, \quad (5)$$

where $\mu_0$ is the magnetic permittivity of vacuum and $\mathbf{h} = (h_x,\ h_y)$. Because the microwave magnetic field of MSL is quasi-static and also because the ferromagnetic film is metallic, we neglected the term involving the electric permittivity on the right-hand side of Eq.(3) and took into account only the term involving the film conductivity $\sigma$.

We also assume that all dynamical variables do not depend on *z* (the quasi-static approach to description of microwave transmission lines), therefore our problem is two-dimensional. The standard real-space methods face serious difficulties [1,30] even in two dimensions because of incompatibility of the length scales in Fig. 1(a) – compare *L* with *w d,* and *s*. To get around this problem and ultimately to accelerate the numerical solution we take advantage of the translational symmetry of the sample in the direction *x*. To this end we Fourier-transform Eqs.(3-5) and (1,2) with respect to *x*.

To implement the Fourier transformation we assume that

$$\mathbf{m}, \mathbf{h} = \int_{-\infty}^{\infty} \mathbf{m}_k, \mathbf{h}_k \exp(-ikx) dx,$$

where

$$\mathbf{m}_k, \mathbf{h}_k = 1/(2\pi) \int_{-\infty}^{\infty} \mathbf{m}, \mathbf{h} \exp(ikx) dx.$$

This procedure results in a system of equations:

$$\partial h_x / \partial y + ikh_y = -\sigma e_z, \quad (6)$$

$$\partial e_z / \partial y = -i\omega\mu_0 (h_x + m_x), \quad (7)$$

$$ke_z = -\omega\mu_0 (h_y + m_y), \quad (8)$$

$$-ikh_x + \partial h_y / \partial y = -\partial m_y / \partial y + ikm_x. \quad (9)$$

Here and at many places below we drop the subscript "*k*" to simplify notations.

We differentiate (6) and substitute (7) into the resulting differentiated equation and make use of (9). This gives:

$$(\partial^2/\partial y^2 - K^2)h_x - K^2 m_x - ik\partial m_y/\partial y = 0\,, \qquad (10)$$

where $K^2 = k^2 + i\sigma\omega\mu_0$.

We seek solutions of the system of equations (2,9,10) for the area inside the film with the spatial variation exp(±*Qy*), as suggested in [9]. After a brief calculation one finds that

$$Q = \sqrt{k^2 - i\mu_0\mu_V\omega\sigma}\ , \ (11)$$

where

$$\mu_V = \left[(\chi+1)^2 - {\chi_a}^2\right]/(\chi+1)\,. \ (12)$$

Accordingly, the complete solution reads

$$\begin{array}{l} h_x = A\exp(Qy) + B\exp(-Qy) \\ h_y = iAC_+\exp(Qy) + iBC_-\exp(-Qy) \end{array}, \ (13)$$

where

$$C_\pm = \left((\chi+1)k \pm \chi_a Q\right)/\left(\chi_a k \pm Q(\chi+1)\right). \ (14)$$

This result is the same as in Ref.[9].

We now need electromagnetic boundary conditions which will relate the electromagnetic fields inside and outside the film. The microwave magnetic field outside the film is given by the same Eq.(10), but for $\sigma = \mathbf{m} = 0$. Let us first consider the area above the film *y*>*L*. From (10) and the condition of vanishing of the microwave magnetic field at *y*=+∞ one easily finds that for *y*>*L*

$$h_y = -i\frac{|k|}{k}h_x\,, \ (15)$$

and at the film surface (*y*=*L*)

$$\frac{|k|}{k}(h_{yi} + m_{yi}) + ih_{xi} = 0\,. \qquad (16)$$

In this expression the subscript "*i*" indicates that these field components are taken at the film surface from *inside* the film. Eq.(16) represents the electromagnetic boundary condition at *y*=*L* which excludes the area *y*>*L* from consideration.

Substitution of Eqs.(13,14) into Eq.(16) allows one to eliminate the unknown integration constant *B*

$$B = AB_0(k)\,, \ (17)$$

where

$$B_0(k) = \exp(2QL)\frac{C_+(1+\chi)|k| + i(k - \chi_a|k|)}{C_-(1+\chi)|k| + i(k - \chi_a|k|)}. \quad (18)$$

A similar boundary condition can be obtained for the area in front of the film ($y$<0). This area contains the strip and the ground plane of the MSL. We model the strip as an infinitely thin sheet of a microwave current $I\mathbf{u}_z$. The linear current density is $j(x)$ (Fig. 1). The width of the sheet along the $x$-axis is $w$; hence $I = \int_{-w/2}^{w/2} j(x)dx$. The sheet is infinite in the $z$ direction to ensure continuity of the current. It is located at a distance $s$ from the film surface $y$=0 (Fig. 1). An electromagnetic boundary condition at the strip reads

$$h_{xk}(y = -s + 0) = h_{xk}(y = -s - 0) - j_k. \quad (19)$$

At a distance $s$+$d$ from the strip the MSL ground plane is located. The ground plane is modelled as a surface of a metal with infinite conductivity ("ideal metal") located at $y$=–$s$–$d$. At the ideal-metal surface $h_{yk}$=0. By solving (10) for $-d-s<y<0$ taking into account (8) together with this boundary condition and the condition (19), one obtains a boundary condition at $y$=0, as follows:

$$(h_{yi} + m_{yi})\coth(|k|(d+s)) - i\frac{|k|}{k}h_{xi} = \frac{\sinh(|k|d)}{\sinh(|k|(d+s))}i\frac{|k|}{k}j_k. \quad (20)$$

Substitution of (13) and (14) into (20) taking into account (17) results in an expression for the integration constant $A$ (see Eq.(13))

$$A = A_0(k)j_k. \quad (21)$$

This concludes solution of the system of equations (2), (6-9). We do no not show the expression for $A_0(k)$ here because it is extremely cumbersome (but easy to derive with some standard software for analytical calculations). We note that for $\sigma$=0, equating the denominator of $A_0(k)$ to zero produces the Damon-Eshbach dispersion relation for the surface magnetostatic wave [31]. This denominator form is typical for solutions of problems of wave and oscillations excitation – zeros of the denominator of an excitation-problem solution correspond to eigen-frequencies and eigen wave numbers of the respective waves or oscillations.

Once the expression for $A$ has been obtained, it is a short exercise to derive an expression for the microwave electric field induced at the strip surface by the dynamical processes in the magnetic film. This expression reads

$$e_{zk}(y = -s) = G_k j_k. \quad (22)$$

where

$$G_k = -\frac{\omega\mu_0}{|k|}\frac{\sinh(|k|d)}{\cosh(|k|(d+s))}\left(\cosh(|k|s) + A_0(k) + A_0(k)B_0(k)\right)j_k. \quad (23)$$

(Here we included the subscript "$k$" for the electric field, in order to recall that all the dynamic quantities above are actually spatial Fourier components of the respective fields.)

The electric field in the real space is obtained by Fourier-transforming Eq.(22) numerically. This is the only numerical step in the solution, provided the Fourier image of the distribution of the microwave current density $j_k$ is assumed to be known (GCD approach). In our previous work [16] we employed the GCD approximation. The main reason for using that method was a very slow numerical code resulting from Eqs.(1,9,10) when the effective exchange field is included in the model.

If the exchange-free limit, the availability of the analytical solution (22) makes the numerical solution very fast, since numerics is needed just to carry out the inverse Fourier transformation of Eq.(22). This allows us to go beyond the approximation of the given current density and, ultimately, to solve the problem of calculation of the electric field self-consistently in the present work. In order to obtain the self-consistent solution, we use the fact that the normal component of the microwave magnetic field $h_y$ should vanish at the surface of the ideal metal of the strip. Then from Eq.(4) it follows that the $z$-component of the electric field $e_z$ induced in the strip by the dynamic magnetisation in the film should be uniformly distributed across the strip width ($e_z(x,y=-s)$=const($x$) for $-w/2<x<w/2$). Without any loss of generality we may set

$$e_z(x, y=-s)=1, \quad -w/2 \le x \le w/2. \ (24)$$

This results in an integral equation

$$1=\int_{-w/2}^{w/2} G_e(x-x')\,j(x')dx', \quad -w/2 \le x \le w/2, \quad (25)$$

where

$$G_e(p)=\int_{-\infty}^{\infty} G_k \exp(-ikp)dk. \quad (26)$$

In this work the Green's function of the electric field $G_e$ is obtained by carrying out the integration in Eq.(26) numerically. We calculate $G_e(p)$ values for $2N$ equidistant points $p_i$ in the range $-w \le p_i \le w$. This transforms Eq.(25) into a vector-matrix equation

$$1=\sum_{i'=1}^{N} \Gamma_{i,i'} j_{i'} \Delta x, \ (27)$$

where $\Gamma_{i,i'}=G_E(x_i-x_{i'})$, $j_i = j(x_i)$ is the unknown distribution of the microwave current density for $N$ equidistant mesh points $x_i$ ($-w/2 \le x_i \le w/2, \quad i=1,2,...N$) and $\Delta x$ is the mesh step. Solving Eq.(27) for $j_i$ using numerical methods of linear algebra results in self-consistent determination of the complex impedance of MSL loaded by the film $Z_r$. The latter quantity is a measure of the microwave magnetic absorption by the film [6]. It may be defined as follows:

$$Z_r = -\frac{U}{I}, \qquad (28)$$

where the linear voltage $U$ (measured in V/m) is the mean value of the total electric field induced at the surface of the strip of MSL:

$$U = \frac{1}{w}\int_{-w/2}^{w/2} e_z(x, y = -s)dx. \quad (29)$$

Given (24), (27) and (29), Eq.(28) reduces to

$$Z_r = \frac{N}{w}\frac{1}{\sum_{i=1}^{N} j_i}, \quad (30)$$

where the vector $j_i$ is the numerical solution of the vector-matrix equation (27).

Once $Z_r$ has been computed, it is a straightforward procedure to calculate the transmission coefficient S21 of the stripline loaded by the ferromagnetic film [10]. The formalism has been explained in detail in [16]. We briefly repeat it here, since it is important for understanding of the mechanism of radiation losses.

We now assume that the film has a finite length $l_s$ along MSL (i.e. in the $y$-direction). The presence of the film on top of the MSL divides MSL into 3 sections – a section covered by the film ("loaded MSL section" or "loaded microstrip" for brevity), a section between the input port of the microstrip fixture and the front edge of the film ("unloaded" microstrip section), and another unloaded section – between the far edge of the film and the output port of the MSL fixture. The unloaded sections of MSL have the same characteristic impedance $Z_c$. The characteristic impedance $Z_f$ of the loaded section is different because of the film presence.

The complex transmission coefficient of the loaded section of MSL reads

$$S21 = \frac{\Gamma^2 - 1}{\Gamma^2 \exp(-\gamma_f l_s) - \exp(\gamma_f l_s)}, \qquad (31)$$

where $\Gamma$ is the complex reflection coefficient from the front edge of the loaded MSL section

$$\Gamma = \frac{Z_f - Z_c}{Z_f + Z_c}, \quad (32)$$

$$\gamma_f = \sqrt{(Z_r(Y_0 + Y_c)} \quad (33)$$

is the complex propagation constant of the loaded microstrip,

$$Z_f = \sqrt{Z_r / (Y_0 + Y_c)}, \quad (34)$$

$Y_0$ is the intrinsic (i.e. in the absence of the sample) parallel capacitive conductance of MSL, and $Y_c$ is the parallel capacitive conductance due to the electric shielding effect [1]. As shown in [16], $Y_c$ is negligible for microstrip lines, therefore we will neglect it below.

## 3. Discussion

### A. Self-consistent approach vs. GCD approach

This algorithm has been implemented as a MathCAD worksheet. The numerical inverse Fourier transform (Eq.(26)) has been carried out using tools for numerical integration built-in MathCAD. The range of integration was $-100\pi / w \leq k \leq 100\pi / w$. To produce the numerical solution, a mesh containing 100 points equidistantly distributed over the strip width was utilized (*N*=100). The linear system of equations (27) was solved using a MathCAD built-in function employing the LU-decomposition method. Since only two steps of the calculation require numerical approaches, the computation is short – a result for 150 values of the applied field *H* is obtained within 30 min.

Fig. 2 shows an example of the self-consistent calculation of the microwave current density distribution across the strip cross-section for two different film thicknesses and two different values of conductivity of the ferromagnetic film. In that graph it is compared with the analytical solution for an unloaded stripline [32]

$$j(x) = 1 / \sqrt{1 - (2x / w)^2}.$$

The spatial Fourier transform of this distribution reads

$$j_k = \frac{w}{2\pi} J_0\left(\frac{kw}{2}\right), \quad (35)$$

where $J_0(p)$ is the Bessel function of first kind of zeroth order.

From this figure one sees that the current distributions are qualitatively similar, the main difference being a much stronger increase in the current density towards the strip edges for the analytical model [32]. One also sees that for non-conducting films the current value needed to obtain 1 Volt/m of microwave electric field at the strip surface is about two times smaller than for ones with conductivity of Permalloy. For the conducting films this value grows with an increase in the film thickness, but for the magneto-insulating ones it remains almost the same (compare the traces for *L*=40 nm and *L*=100nm in Fig. 2).

The similarity of the current density distributions in Fig. 2 translates into similarity of the applied field dependences of the stripline linear impedance (Fig. 3). One clearly sees that the results of the self-consistent calculation and of the GCD model employing Eq.(35) to calculate $Z_r$ are quite close. This justifies the use of the GCD approach in our previous work [16].

It is worth noticing excellent agreement of the result in Fig.(3) obtained in the framework of the GCD model with a respective calculation with the dipole-exchange numerical code [16] (not shown). The amplitude of the resonance peaks in Fig. 3 and the off-resonance level of $Z_r$ are the same in both cases, the only difference between the two calculations being the presence of a higher-order standing-spin-wave peak in the dipole-exchange model (see e.g. Fig. (2) in [16]). The peak appears for large sample conductivity values and is due to the eddy-current contribution to excitation of magnetization dynamics by stripline transducers [10].

Also, like in [16], the obtained value of Im($Z_r$) off resonance is in agreement with a result of calculation with the known analytical formula for the linear inductance for microstrip lines (see [16] fore more detail of this property of $Z_r$ for non-conducting films). For instance, in Fig. 3(b) the off-resonance Im($Z_r$)=327 Ohm/cm (as measured for $H$=0), and the independently calculated linear inductance for this microstrip geometry is 332 Ohm/cm.

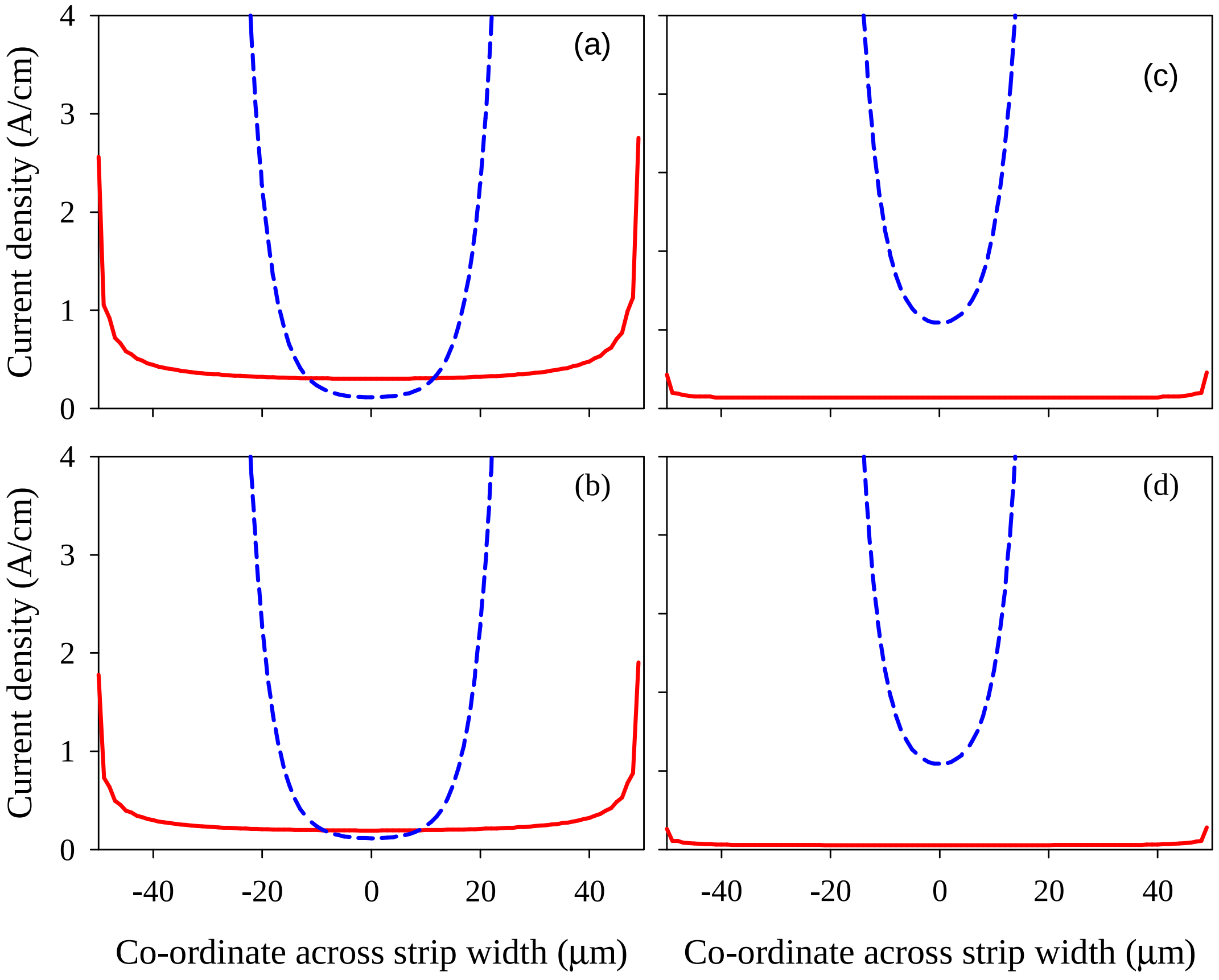

Fig. 2. Distributions of microwave current density across the width of the strip. Solid lines: results of the self-consistent calculation. Dashed lines: calculated with the analytical formula (35). For all panels, film thickness $L$=40nm. (a) and (b): strip width $w$=100 micron. (c) and (d): $w$=1500 micron. (a) and (c): film conductivity $\sigma$=4.5x10$^6$ S/m. (b) and (d) $\sigma$=0.

Microwave frequency is 9.5 GHz, film saturation magnetization ($4pM_0$) is 10 kG, gyromagnetic ratio is 2.8 MHz/Oe, Gilbert magnetic loss parameter is 0.008.

### B. Radiation losses

Let us now discuss the important aspect of coupling of magnetization dynamics in the film to the electromagnetic field of the stripline transducer. As mentioned in the introduction, non-negligible coupling results in additional contribution to the resonance linewidth due to "radiation losses" [27]. Below we will carry out calculations with the constructed model in order to elucidate importance of the radiation losses effect and its potential correlation with the eddy-current shielding effect [10]. Above we have established that the GCD model and the self-consistent one deliver practically equivalent results. Therefore we will use numerical results obtained with the given-current density model for this discussion. A big advantage of the GCD method is that software based on it is very fast. Given that

$$e_z(x, y=-s) = \int_{-\infty}^{\infty} e_{zk}(x, y=-s)(-ikx)dk \; , (36)$$

in the framework of the GCD approach Eq.(29) reduces to

$$U = \frac{w}{2\pi}\int_{-\infty}^{\infty} e_{zk}(x, y=-s)\frac{\sin(kw/2)}{kw/2} J_0\left(\frac{kw}{2}\right)dk \,. \quad (37)$$

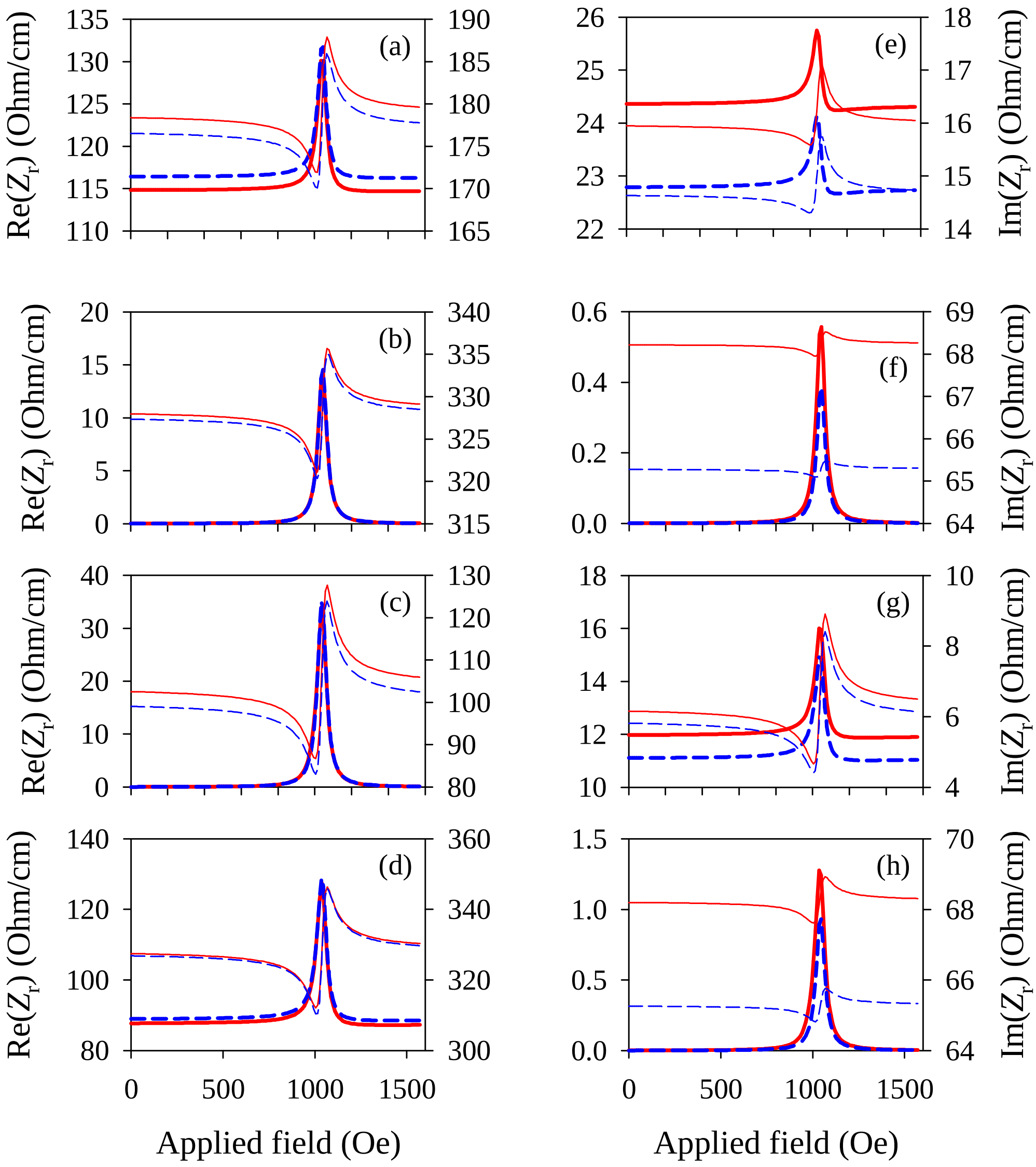

Fig. 3. Comparison of results of the self-consistent calculation (thick lines) with results obtained with the Given Current Density model (thin lines). Solid lines: real parts of impedance $Z_r$. (left-hand vertical axes). Dashed lines: imaginary parts of the impedance (right-hand vertical axes.) Left-hand column: strip width $w$=100 micron. Right-hand column: $w$=1500 micron. (a),(e), (c), and (g): film conductivity $\sigma$=4.5x10$^6$ S/m. (b), (f), (d), and (h): $\sigma$=0. (a),(b), (e) and (f): film thickness $L$=40 nm. (c), (d), (g) and (h): $L$=100 nm. All other parameters of calculation are the same as for Fig. 2.

Taking the integral in (37) is the only numerical step of this calculation. Therefore, the corresponding MathCAD worksheet is very quick – it takes about 10 seconds to complete a program run for 150 values of the applied field.

Fig. 4 compares results of calculation of Re(S21) and Re($Z_r$) [33]. One sees a noticeable difference in linewidths for the two peaks. We fit complex S21 and $Z_r$ with a complex Lorentzian

$$F(H) = D_0 + \frac{D_1}{H - D_2} \,, \quad (38)$$

where $D_0$, $D_1$ and $D_2$ are allowed to take complex values. The fits reveal that the resonance linewidths (given by Im($D_2$)) are different for S21 and $Z_r$ – 45 Oe and 31 Oe respectively.

Fig. 5(a) demonstrates the extracted resonance linewidth dependence on the strip width. Clearly, the difference in the linewidths for the two quantities decreases with an increase in $w$. From this figure one also sees that an increase in the thickness $s$ of the spacer between the stripline and the film also decreases the difference.

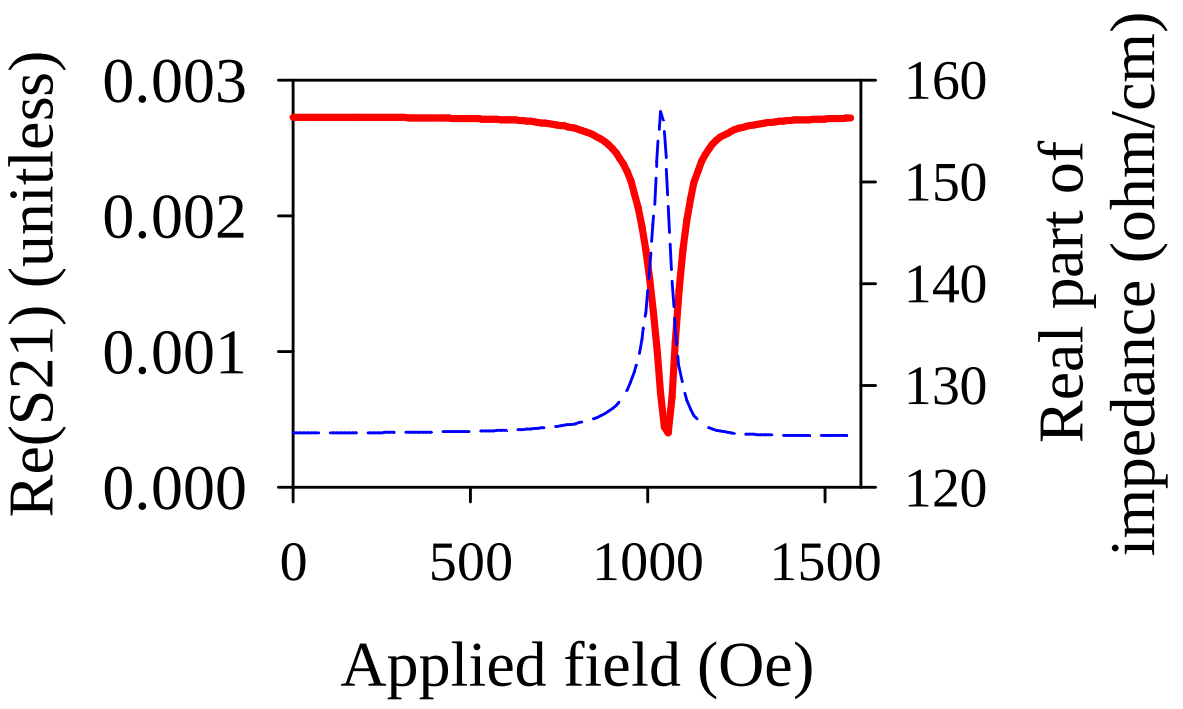

Fig. 4. Comparison of Re($Z_r$) (dashed line, right-hand axis) and Re(S21) (solid line. left-hand axis) traces. Film thickness is 60 nm, strip width is 100 micron, conductivity $\sigma$=4.5x$10^6$ S/m. Film length along the strip is 7 mm. All other parameters of calculation are the same as for Fig. 2.

Let us now discuss the origin of this behaviour. The relation between S12 and $Z_r$ is given by Eqs.(31-34). From these formulas it becomes clear that the linewidth of the peak in $Z_r(H)$ represents a kind of "internal" linewidth for the ferromagnetic resonance in the film if we extend the notion of the "external quality factor" [28] onto the linewidth parameter. Similarly, the linewidth of the resonance peak in the S21($H$) dependence is then the "external linewidth", since it includes the strength of the coupling of the sample to the input and the output ports of the probing stripline fixture. Actually, Eqs.(31-34) take into account how strongly the section of the microstrip line covered by the sample ("loaded section") couples to the sections of the stripline in front and behind the sample ("unloaded sections"). Important is the mismatch in the characteristic impedances of the loaded section $Z_f$ and the unloaded ones $Z_c$ (see

Eq.(32)). As follows from Fig. 2, for magneto-insulating films, off resonance, $Z_f = Z_c$. Thus, a perfect impedance match takes place. (In real experiments it will be some extra mismatch due to the dielectric constants of the film and the film substrates. These constants are not taken into account in the present theory.) The perfect impedance match means strong coupling and, consequently, significant broadening of the external resonance line. On the contrary, for the films with large conductivity of metals, due to large contribution of microwave eddy currents in the films to $Z_r$ off resonance, there is an impedance mismatch for any value of the applied field. Consequently, the internal and the external resonance linewidths are closer to each other.

Physically, an impedance mismatch between the loaded and the unloaded sections of the stripline implies that a wave of electric field $e_z$ (Eq.(36)) induced in the strip by the magnetization dynamics in the film gets partly trapped within the loaded section. This happens because the mismatch leads to a non-vanishing reflection from the edges of this section (i.e. from the stripline cross-sections along the *y*-*z* plane corresponding to the edges of the film sample) and, consequently, to internal reflection of the wave incident on the section edges from inside the loaded section. The energy of the microwave electric field is a part of the total energy of the ferromagnetic resonance in the sample. The larger the mismatch, the more energy of the electric field of dynamic magnetization is trapped below the film, the less resonance energy escapes into the two unloaded sections of the stripline. Consequently, with an increase in the mismatch, the loaded quality factor of the ferromagnetic resonance increases which ultimately leads to a smaller external resonance linewidth.

One more important observation from Fig.(5a) is that the linewidth of the resonance peak in $Z_r$ (thick dashed line) varies noticeably as a function of *w* for *s*=0. One also sees that this trace converges with the respective trace for *s*=33 μm (thin dashed line) for large values of *w*. The two facts suggest that the radiation losses do not depend solely on coupling between the loaded and unloaded stripline sections. There exists one more contribution to the total radiation losses and this contribution reduces with an increase in *s*. This contribution is coupling of the microwave field of the loaded section of the stripline to the magnetization dynamics in the film. The electric field $e_z$ of dynamic magnetization represents an evanescent wave outside the film – it decays exponentially with the distance from the film surface. Therefore, the farther the stripline is located from the film surface, the smaller the microwave voltage induced by the dynamic magnetization across the length of the loaded section of the stripline is. This implies that increasing *s* removes one more contribution to the radiation losses – the one related to coupling of the film to the stripline beneath it.

Furthermore, for the large separation *s*=33 micron the external linewidth (as extracted from the S21 traces) becomes very close to the one extracted from the $Z_r$ data (compare the two thin lines in Fig. 5(a)). This suggests that the coupling of the ferromagnetic resonance to the ports of the stripline fixture is a two-step process. The first step is coupling of the electric field of the dynamic magnetization to the strip located below the film. The second step of the process is coupling of the loaded section of MSL to the two unloaded ones via the microwave voltage induced across the loaded section.

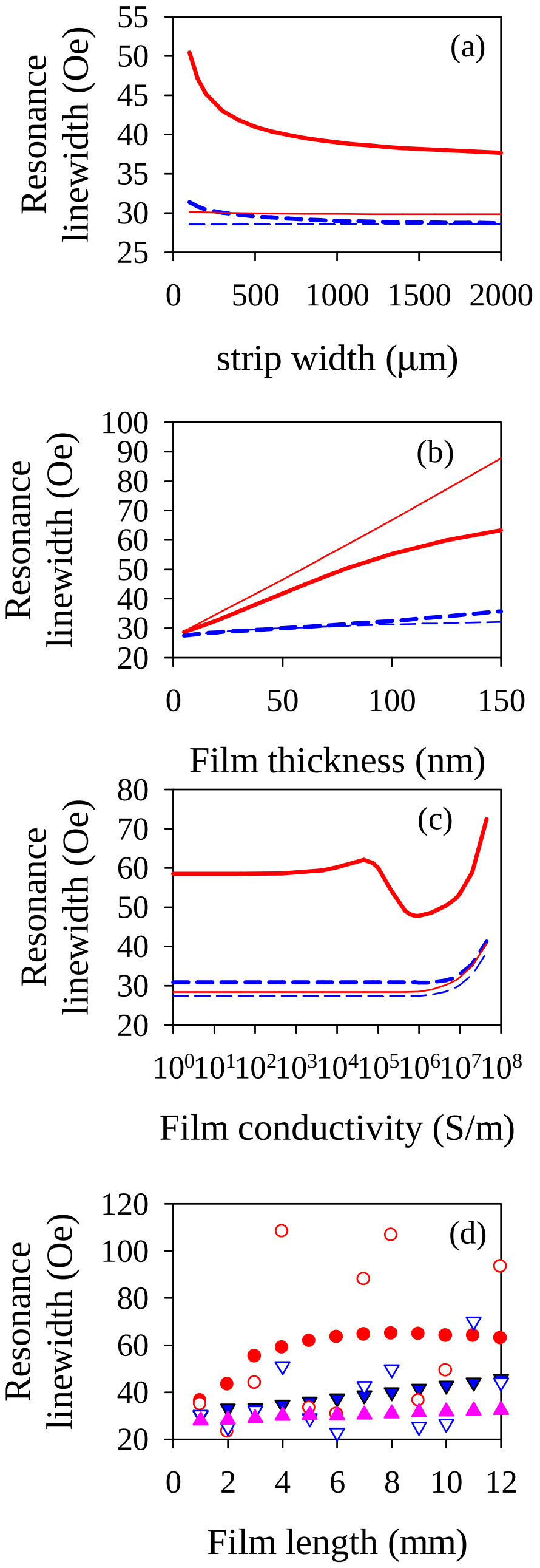

Fig. 5. (a) Resonance linewidth as a function of the strip width. Thick lines: no spacer between the film and the stripline (but no electric contact between the two). Thin lines: spacer thickness $s$=33 micron. Solid lines lines: S21; dashed lines: $Z_r$. (b) Resonance linewidth vs. film thickness. Thick lines: $\sigma$=4.5x10$^6$ S/m; thin lines: $\sigma$=0. Solid lines: S21; dashed lines: $Z_r$. Strip width is 100 micron. $s$=0 for all plots in this panel. (c) Resonance linewidth as a function of film conductivity. Solid lines: S21; dashed lines: $Z_r$. Thin lines: $s$=33 micron; thick lines: $s$=0. Strip width is 100 micron. (d) Linewidth of the peak in S21 traces as a function of the sample size $l_s$ along the microstrip. Open circles: $\sigma$=0; $L$=150nm; filled circles: the same, but $\sigma$=4.5x10$^6$ S/m. Open triangles down: $\sigma$=0; $L$=40nm; filled triangles down: the same, but $\sigma$=4.5x10$^6$ S/m. Filled triangles up: the same as filled triangles down, but the strip thickness is now finite and is equal to 16 micron. For all graphs in (d) $w$=100 micron and $s$=0. All other parameters of the calculation are the same as for Fig. 2. $l_s$ for (a)-(c) is 7 mm.

If the first step is inefficient, the overall coupling strength will be small, independently from the efficiency of the second step. This explains why the difference in the linewidths for the peaks in S21 and $Z_r$ decreases with an increase in *s*. The increase in *s* leads to a decrease in the efficiency of the first step of the coupling process and hence to a decrease in the total coupling strength.

The analysis above suggests that in order to measure the unloaded (or intrinsic) resonance linewidth of a sample very precisely, it is important to keep the strength of coupling for the first step of the coupling process weak. The latter is achieved by lifting the sample from the film surface. Also, the coupling is weaker for wider striplines. This is related to smaller current density in a wider stripline for the same $Z_c$ and the same microwave voltage applied to the input port of the line. As $e_z$ scales as $j$ (see Eq.(22)), increasing $w$ decreases $e_z$ and hence decreases the efficiency of the first step of the coupling process. For conducting samples, the eddy-current contribution to the FMR dynamics helps to further decrease the coupling, since this contribution increases with an increase in $w$ [16]. Stronger microwave shielding by the eddy currents in the film for larger $w$ values leads to a stronger decrease in $Z_r$ off resonance and hence to stronger impedance mismatch leading to a decrease in the strength of the second stage of coupling.

Unfortunately, both measures – lifting the sample and increasing the stripline width - will also result in a decrease in the strength of the FMR response – the height of the peak in S21. On the other hand, the peak height grows with an increase in the film thickness $L$ (see Fig. 3). Fig. (5b) demonstrates the effect of the film thickness on the resonance linewidth. One sees that the resonance linewidth grows with an increase in $L$. The effect is much stronger for S21 than for $Z_r$. Hence, the second step of the coupling process contributes more to this effect. Interestingly, for S21 the linewidth broadening is larger for magneto-insulating films (thin solid line) than for ones with conductivity of metals (thick solid line). For $Z_r$ it is other way around – the linewidth broadening is stronger for conducting films (compare the two dashed lines).

This is due to contribution of eddy current losses to the intrinsic linewidth for thicker films. For thinner films the two dashed lines overlap with graphic accuracy. The latter confirms the known fact that contribution of eddy-current losses to the intrinsic FMR *linewidth* for metallic films with small thicknesses is negligible. It also suggests that on resonance the contribution of eddy currents to $e_z$ is the same as off-resonance. Hence, the field which is responsible for the first step of the coupling process is the electric field of dynamic magnetization. In other words, the contribution to the first step of the coupling process by the electric field induced by the Oersted field of the eddy current in the film through Faraday induction is negligible.

The fact that the linewidth broadening for the peak in S21 is larger for insulating films confirms our conclusion above that shielding by eddy currents reduces the strength of coupling for the second step of the coupling process. A calculated dependence of the resonance linewidth on film conductivity is shown in Fig. 5 (c ). The main observations from this figure are that in the 10-GHz frequency range, only the large conductivity of metals matters. Conductivity values below $10^4$ S/m do not affect the broadband ferromagnetic resonance linewidth. Noteworthy is a significant drop in the peak width seen in the S21 traces between $10^4$ and $10^5$ S/m. This is due to the onset of the above-discussed eddy-current induced decoupling of the ferromagnetic resonance in the film from the environment. Also, noteworthy is simultaneous growth of all four curves in Fig. 5(c ) for $\sigma > 10^6$. As this effect is also present for a strongly decoupled resonance ($s$=33 μm, thin lines), it is not due to the radiation losses, but to an increase in the eddy-current contribution to intrinsic FMR losses for larger film conductivities.

Fig. 5(d) demonstrates the effect of the sample length along the stripline $l_s$ on the width of the external resonance line. This parameter enters Eq.(31) only, hence it affects the second step of the coupling process only. One sees that the impacts of $l_s$ on magneto-insulating and conducting films are quite different. For the insulating films the model shows strong periodic variation of the linewidth, but for the films with large conductivity the dependence is smooth and the change in the linewidth across the displayed range of film lengths is much smaller. The quasi-periodic character may be explained taking into account that $l_s$ enters the arguments of the exponential functions in Eq.(31). Given that $Y_c$ is negligible and $Y_0$ is an imaginary quantity, the propagation constant $\gamma_f$ is imaginary for $\sigma$=0 and off resonance. A small real part is added to it on resonance (compare the scales of the right-hand and left-hand vertical axes in Fig. 3(b)). However, because on resonance the imaginary part of $\gamma_f l_s$ still dominates, $\exp(\gamma_f l_s)$ remains a (quasi)periodical function. In Fig. 5(d) this character is seen as a significant scatter of data points for $\sigma$=0. For the films with large conductivity of metals, $\gamma_f$ is essentially complex both on and off resonance (see e.g. Figs. 3(a) or (e)). As a result, the S21 dependence on $\gamma_f l_s$ does not have a pronounced quasi-periodic character.

So far we have treated the stripline as infinitely thin in the *y*-direction. The GCD model allows one to consider the effect of the strip thickness, provided the distribution of the microwave current density *j*(*x*,*y*) is known. For simplicity, let us assume that it is uniform across the strip thickness $t_s$, Then, $Z_r$ may be estimated as

$$Z_r = -\frac{\frac{1}{S}\int_S G_e(y,y',x,x')\, j(y',x')dy'dx'}{\int_S j(y,x)ds}, \quad (39)$$

where $S = wt_s$ is the strip cross-section area. The Green's function $G_e(y,y',x,x')$ is obtained from (17,23) by considering the electric field of dynamic magnetisation at the point (*y*,*x*) of the strip cross-section, provided that the dynamic magnetisation is driven by an infinitely thin wire of current at the point (*y'*,*x'*), also belonging to the same cross-section.

An example of numerical calculation by using (39) is shown in the same Fig. 5(d). One sees that inclusion of the final strip thickness drastically reduces the external linewidth broadening. The effect is very similar to lifting an infinitely thin strip by some distance *s* on the order of $t_s/2$ from the film surface. Thus, the finite thickness of real striplines is a very important factor that naturally reduces coupling of the film to the environment in real experiments.

## 4. Conclusion

In this work we constructed a quasi-analytical self-consistent model of strip-line based broadband ferromagnetic resonance experiment. With this model we studied the contribution of radiation losses to the ferromagnetic resonance linewidth. We found that for films with large conductivity of metals the radiation losses contribution is significantly smaller. This is because of impedance mismatch due to excitation of microwave eddy currents in these materials. We also show that the radiation losses drop with an increase in the stripline width and when the sample sits at some elevation from the stripline surface. Furthermore, the radiation losses contribution is larger for thicker films.

Two consecutive steps of coupling of the ferromagnetic resonance to the environment leading to the radiation losses have been identified. The first one is coupling of the dynamic magnetization to the stripline section on top of which the film sits ("loaded" section). This coupling proceeds via the microwave electric field associated with magnetization precession. The second step of the process is coupling of the microwave electric voltage induced in the loaded section to unloaded sections of the stripline which join the loaded section to the input and the output ports of the stripline fixture.

The impedance mismatch affects the second step of the coupling process. The stripline width and its distance from the film surface are important for the first step. By minimizing the coupling strength for the first step of the process it is possible to significantly reduce total radiation losses.

Thus, in order to eliminate the measurement artefact of radiation losses in real broadband stripline ferromagnetic resonance experiments one needs to employ wide striplines and introduce a spacer between the film and the sample surface.

**Acknowledgment**

Financial support by the Australian Research Council, the University of Western Australia (UWA) and the UWA's Faculty of Science is acknowledged.